\documentclass[article, twocolumn,
   aps, pra,
  amsmath,amssymb,
  longbibliography,
  ]{revtex4-1}
\usepackage{graphicx,color}
\usepackage{amsmath}
\usepackage{natbib}
\usepackage{epsfig}
\begin{document}

\title{Freezing density scaling of transport coefficients in the Weeks-Chandler-Andersen fluid}

\author{S. A. Khrapak}\email{Sergey.Khrapak@gmx.de}
\affiliation{Joint Institute for High Temperatures, Russian Academy of Sciences, 125412 Moscow, Russia}
\author{A. G. Khrapak}
\affiliation{Joint Institute for High Temperatures, Russian Academy of Sciences, 125412 Moscow, Russia}

\begin{abstract}
It is shown that the transport coefficients (self-diffusion, shear viscosity, and thermal
conductivity) of the Weeks-Chandler-Anderson (WCA) fluid along isotherms exhibit a freezing density scaling (FDS). The functional form of this FDS is essentially the same or closely related to those in the Lennard-Jones fluid, hard-sphere fluid, and some liquefied noble gases. This proves that this FDS represents a quasi-universal corresponding state principle for simple classical fluids with steep interactions. Some related aspects such as Stokes-Einstein relation without a hydrodynamic diameter and gas-to-liquid dynamical crossover are briefly discussed. Simple fitting formula for the transport coefficients of the dense WCA fluid are suggested.       
\end{abstract}

\date{\today}

\maketitle

\section{Introduction}

Understanding transport properties of fluids is crucial in fields such as chemical physics and engineering, condensed matter, materials science, and environmental science. Despite certain progress~\cite{FrenkelBook,BalucaniBook,MarchBook,HansenBook} our understanding remains rather limited and fragmented. It is unlikely that a general theory can be developed, allowing to describe the actual transport properties of fluids with varying molecular composition, interactions, location on the phase diagram and other conditions. Even for simple fluids such a task seems unrealistic. For this reason, various scaling relationships and empirical correlations still play a very important role~\cite{RosenfeldPRA1977,DzugutovNature1996,DyreJCP2018,BellJPCB2019,KhrapakPhysRep2024}.  

It has been recently discovered that appropriately reduced self-diffusion, shear viscosity, and thermal conductivity coefficients of the Lennard-Jones (LJ) fluid along isotherms exhibit quasi-universal scaling on the density divided by its value at the freezing point ~\cite{KhrapakPRE04_2021,KhrapakJPCL2022,KhrapakJCP2022_1,
HeyesJCP2023}. 
This freezing density scaling (FDS) implies that the transport coefficients are functions of a single variable $\rho/\rho_{\rm fr}$, where $\rho$ is the number density and $\rho_{\rm fr}$ is its value at the freezing point. Since the dependence of $\rho_{\rm fr}$ on temperature is relatively well known for the LJ system, FDS represents a very convenient corresponding states principle to estimate transport properties of the LJ fluid.

Originally considered as a useful empirical observation~\cite{KhrapakPRE04_2021}, the FDS has been later discussed in the context of   
quasi-universal excess entropy scaling and isomorph theory~\cite{KhrapakJPCL2022,KhrapakJCP2022_1,HeyesJCP2023}.
It occurs, however, that FDS holds even at quite low densities where neither the original form of the excess entropy scaling (predicting exponential dependence of transport coefficients on the excess entropy~\cite{RosenfeldPRA1977,RosenfeldJPCM1999}) nor the isomorph theory are expected to work. The applicability domain of the FDS approach covers almost the entire gas-liquid-fluid region in the LJ system phase diagram (probably excluding very low densities, deep gas-liquid coexistence region, and the vicinity of the critical point), including metastable
regions of superheated and supercooled liquids and supersaturated vapour~\cite{KhrapakPRE12_2023}.

Remarkably, the functional form of the FDS scaling in the LJ fluid  is similar (although not identical) to that in the hard-sphere (HS) fluid~\cite{KhrapakPRE04_2021}. It also applies reasonably well to liquefied noble gases~\cite{KhrapakJCP2022_1}, where the LJ potential can serve only as a rough approximation. Thus, it can be expected that FDS is not a special property of the LJ fluid, but applies to a wider class of fluids. The purpose of this paper is to validate this hypothesis using the Weeks-Chandler-Anderson (WCA) model. In particular, we demonstrate that the reduced self-diffusion, viscosity, and thermal conductivity coefficients of the WCA fluid exhibit {\it the same} FDS scaling as the LJ fluid. This finding proves that the FDS approach is a very useful practical tool to estimate the transport coefficients of various simple fluids in a rather wide parameter regime.  

\section{Formulation}

We consider a classical system of $N$ particles of mass $m$ confined to the volume $V$ (so that $\rho = N/V$) and interacting via the WCA repulsive potential~\cite{WeeksJCP1971}: 
\begin{equation}
\phi(r)= 4\epsilon\left[\left(\frac{\sigma}{r}\right)^{12}-\left(\frac{\sigma}{r}\right)^6\right]+\epsilon \quad\quad {\rm for} \quad r<r_c 
\end{equation}
and zero otherwise. Here $\epsilon$ and $\sigma$ are the energy and length scales, and $r_c=2^{1/6}\sigma$ is the cut-off distance. The conventional reduced units for temperature and density are $T^*=k_{\rm B}T/\epsilon$ and $\rho^*=\rho\sigma^3$, respectively.

The WCA potential corresponds to the LJ potential, truncated and shifted at the location of its minimum. It was originally used as a reference repulsive potential when separating the LJ potential into the repulsive and attractive parts~\cite{WeeksJCP1971}. Nowadays, the WCA system is often used as one of the models of simple fluids~\cite{KuijperJCP1990,HeyesJCP2006,NasrabadJCP2008,AhmedPRE2009,
KhrapakJCP2011_3,BellPNAS2019,AttiaJCP2022,DyreJCP2023}. 

Since the WCA potential is purely repulsive, WCA system does not exhibit gas-liquid phase transition and gas-liquid coexistence region. There are no critical or triple points in this model. Its phase diagram is thus very different from that of the LJ system. Similarly to the hard-sphere system only the fluid-solid phase transition is present. Thus, identical values of $\rho/\rho_{\rm fr}$ can be associated with very different phase states for the LJ and WCA systems. 
%It is far from obvious that similar scalings can operate in both systems simultaneously.  

The transport coefficients of the WCA fluid are taken from Ref.~\cite{NasrabadJCP2008}, where equilibrium molecular dynamics simulations have been used to compute thermodynamic, structural and dynamical quantities. The data cover relatively wide region of the phase diagram: Five isotherms with $T^*=0.5$, $1.0$, $2.0$, $4.0$, $6.0$ and the density in the range $0.3\leq \rho^*\leq 1.0$. The tabulated data are found in the Supplementary Material of Ref.~\cite{NasrabadJCP2008}. The transport coefficients of the LJ and WCA fluids are usually reduced using $\sigma$ as a unit of distance, $\sqrt{\epsilon/m}$ as a unit of velocity, and $\sigma/\sqrt{\epsilon/m}$ as a unit of time. To compare transport properties in different systems it is more convenient to use a universal macroscopic normalization (sometimes referred to as Rosenfeld's normalization~\cite{RosenfeldJPCM1999}) with mean interparticle separation $\Delta = \rho^{-1/3}$ as a unit of distance, thermal velocity $v_{\rm T}=\sqrt{k_{\rm B}T/m}$ as a unit of velocity, and $\Delta/v_{\rm T}$ as a unit of time. The reduced self-diffusion ($D$), viscosity ($\eta$), and thermal conductivity ($\lambda$) coefficients are
\begin{equation}\label{Rosenfeld}
D_{\rm R}  =  D\frac{\rho^{1/3}}{v_{\rm T}} , \quad
\eta_{\rm R}  =  \eta \frac{\rho^{-2/3}}{m v_{\rm T}}, \quad \lambda_{\rm R}  =  \lambda \frac{\rho^{-2/3}}{v_{\rm T}}, 
\end{equation}         
Here the subscript $R$ denotes Rosenfeld normalization. We have recalculated the original data presented in Ref.~\cite{NasrabadJCP2008} to get $D_{\rm R}$, $\eta_{\rm R}$, and $\lambda_{\rm R}$.

The data on the fluid-solid coexistence of the WCA model can be found for example in Refs.~\cite{AhmedPRE2009,MirzaeiniaJCP2017,DyreJCP2023,AttiaJCP2022}. In a recent publication~\cite{AttiaJCP2022}, the points on the fluid and solid boundaries of the fluid-solid coexistence have been determined by numerical integration of the
Clausius-Clapeyron relation in a very wide range of $T^*$. For the temperature range $0.5\lesssim T^*\lesssim 6$, relevant for the present investigation, we use a fourth order polynomial to fit the tabulated data for the liquidus. This yields
\begin{align*}
\rho^*_{\rm fr}&\simeq 0.77622+0.2346T^*-0.05393(T^*)^2\\
               &+0.00789(T^*)^3-4.56519\times 10^{-4}(T^*)^4.
\end{align*}
This expression is used to evaluate the freezing density and the FDS scaling parameter ${\mathcal R}=\rho/\rho_{\rm fr}$.

\section{Results}

\subsection{Stokes-Einstein relation}

First, let us verify that the Stokes-Einstein (SE) relation between the self-diffusion and viscosity coefficients is satisfied in the WCA fluid. We chose the form known as SE relation without a hydrodynamic diameter~\cite{CostigliolaJCP2019}  
\begin{equation}\label{SE}
D\eta\left(\frac{\Delta}{k_{\rm B}T}\right)\equiv D_{\rm R}\eta_{\rm R}=\alpha_{\rm SE},
\end{equation} 
where $\alpha_{\rm SE}$ is a numerical coefficient. In this version of the SE relation the effective hydrodynamic diameter is essentially replaced by the mean interparticle separation. There exist different theoretical approaches explaining why it should be approximately so~\cite{LiJCP1955,ZwanzigJCP1983,Balucani1990}. The constant $\alpha_{\rm SE}$ is weakly system-dependent~\cite{KhrapakMolPhys2019,KhrapakPRE10_2021}. For simple fluids, the general tendency is that it increases from $\alpha_{\rm SE}\simeq 0.14$ for plasma-related systems with soft (Coulomb-like) interparticle interaction to $\alpha_{\rm SE}\simeq 0.17$ for extremely steep hard-sphere interactions~\cite{KhrapakPRE10_2021}.    

Numerous confirmations of the applicability of the SE relation without the hydrodynamic diameter to dense simple fluids have been reported, including charged particles in the plasma-related context~\cite{DaligaultPRE2014,KhrapakAIPAdv2018,
KhrapakMolecules12_2021,KhrapakPRE10_2021}, soft (inverse power) repulsive particle fluid~\cite{HeyesPCCP2007}, the LJ fluid~\cite{CostigliolaJCP2019,OhtoriPRE2015,OhtoriPRE2017,
KhrapakPRE10_2021}, and the HS fluid~\cite{OhtoriJCP2018,Pieprzyk2019,KhrapakPRE10_2021}.
Several important non-spherical molecular liquids have been examined using numerical simulations in Ref.~\cite{OhtoriChemLett2020} and  applicability  of the SE relation (\ref{SE}) has also been confirmed. This indicates that the applicability regime is apparently wider than simple atomic systems with spherically symmetric interactions. Among recent demonstrations of the applicability of SE relation in the form of Eq.~(\ref{SE}) to real liquids we mention iron at conditions of planetary cores~\cite{LiJCP2021}, dense supercritical methane (at least for the most state points investigated)~\cite{Ranieri2021,KhrapakJMolLiq2022}, silicon melt at high temperatures~\cite{Luo2022}, and water~\cite{KhrapakJCP2023,KhrapakJPCB2024}.   

In view of previous investigations of the SE relation in the WCA fluid~\cite{CappelezzoJCP2007,OhtoriJCP2018}, there should be no doubt that it applies in this case. The main questions are related to the applicability regime and the value of the constant $\alpha_{\rm SE}$. 

\begin{figure}
\includegraphics[width=8.5cm]{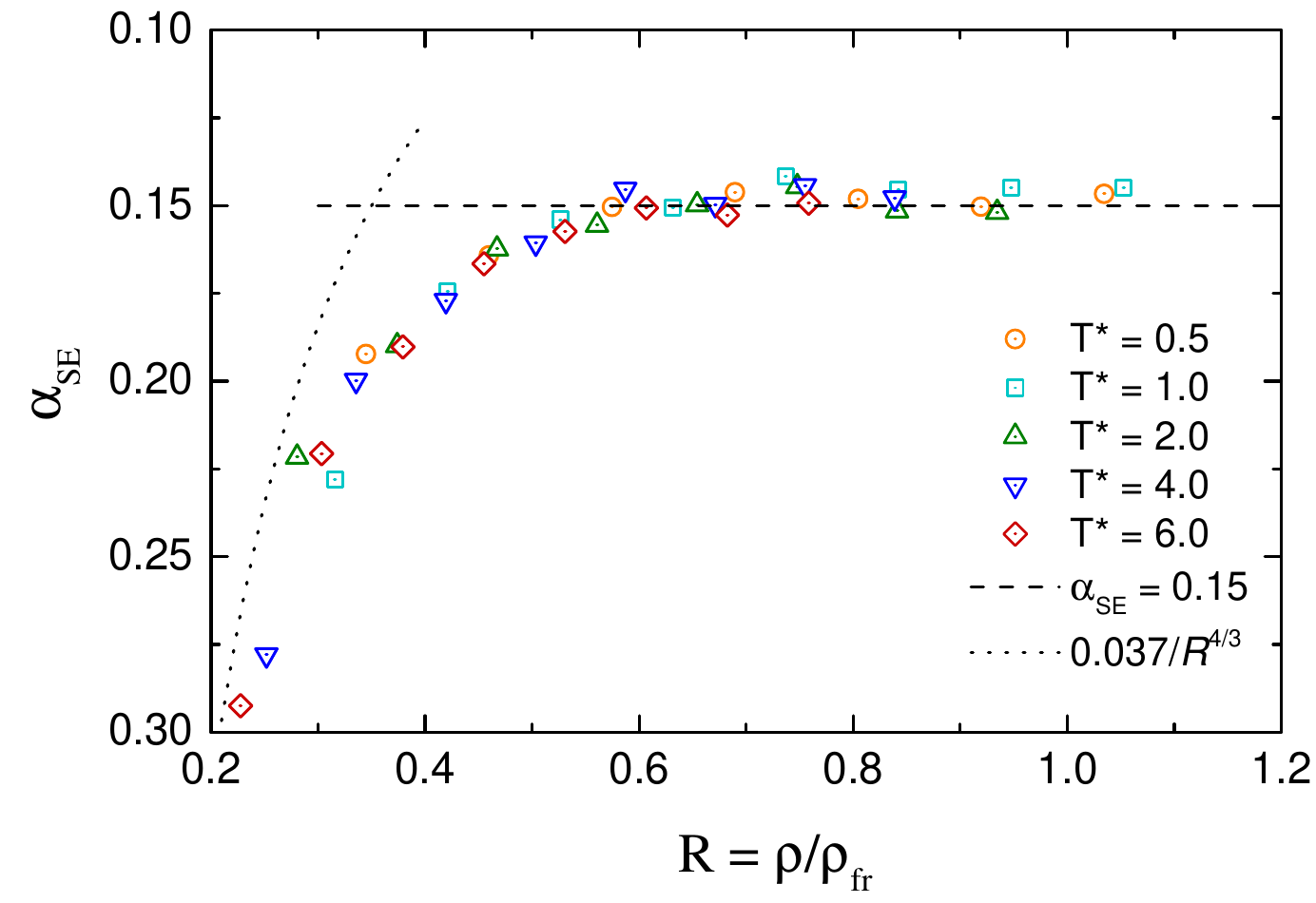}
\caption{(Color online) Stokes-Einstein coefficient $\alpha_{\rm SE}$ in the WCA fluid as a function of the FDS scaling parameter ${\mathcal R}=\rho/\rho_{\rm fr}$. The symbols correspond to the numerical results from Ref.~\cite{NasrabadJCP2008}. As the fluid-solid phase transition is approached (${\mathcal R}=1$) the coefficients saturates at a constant value $\simeq 0.15$, shown by the dashed line. The dotted curve corresponds to dilute gas asymptote $\sim {\mathcal R}^{-4/3}$. The vertical axis is reversed to highlight the accuracy of the SE relation at high densities. }
\label{FigSE}
\end{figure}

\begin{figure*}
\includegraphics[width=12cm]{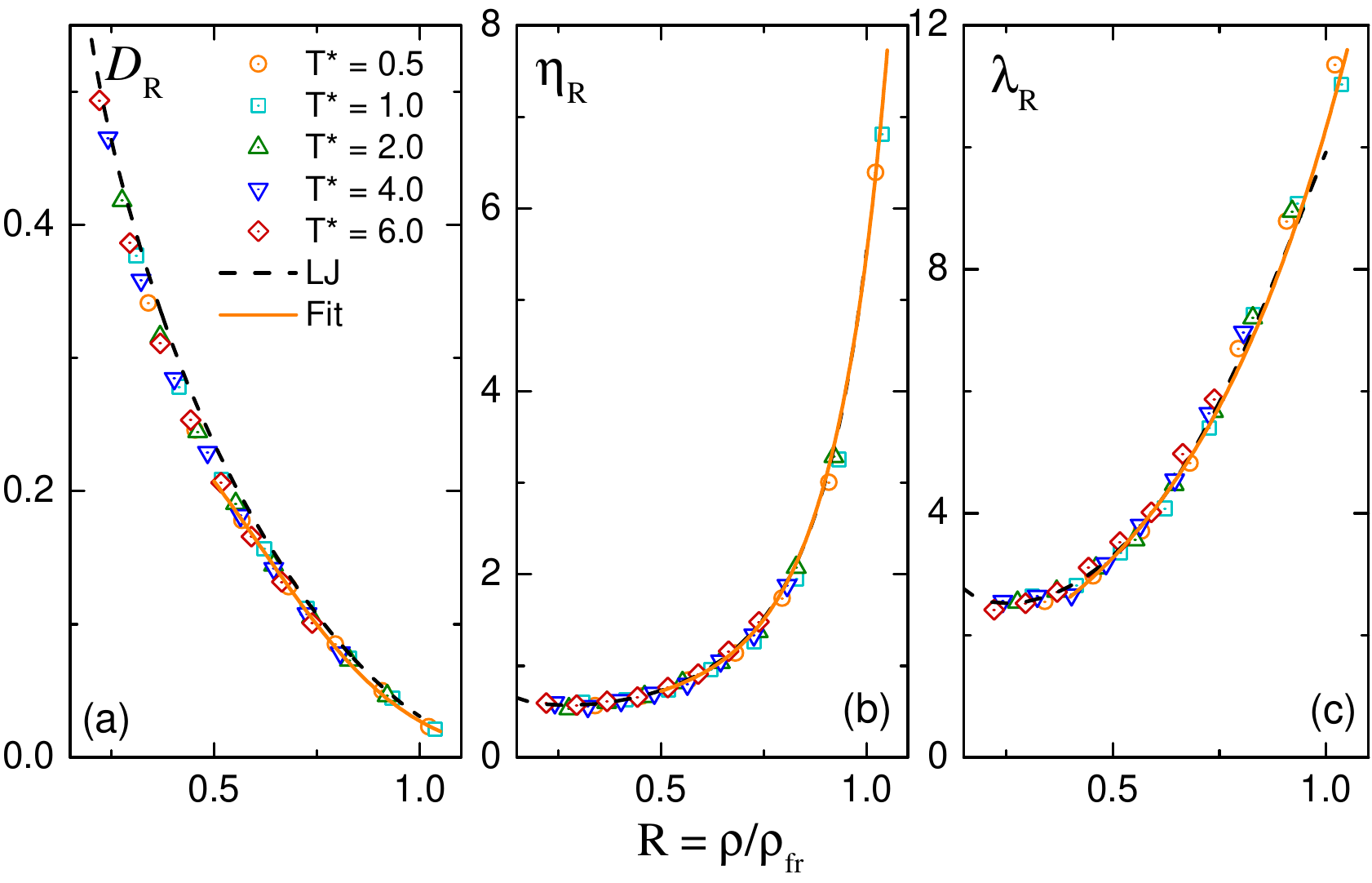}
\caption{(Color online) Reduced self-diffusion (a), viscosity (b), and thermal conductivity (c) coefficients of the WCA fluid versus the FDS scaling parameter ${\mathcal R}=\rho/\rho_{\rm fr}$. Results for five isotherms: $T^*=0.5$, $1.0$, $2.0$, $4.0$, $6.0$ denoted by symbols correspond to the numerical results from Ref.~\cite{NasrabadJCP2008}. The dashed curves correspond to the LJ model FDS evaluated using the modified excess entropy scaling of Ref.~\cite{BellJPCB2019} for an isotherm $T^*=2.0$. The solid curves are the exponential fits of Eqs.~(\ref{DR})-(\ref{lambdaR}) applicable in the dense fluid regime ${\mathcal R}\gtrsim 0.6$. }
\label{FigTransport}
\end{figure*} 

The dependence of $\alpha_{\rm SE}$ on the FDS scaling parameter ${\mathcal R}$ is shown in Fig.~\ref{FigSE}. We observe that at low densities $\alpha_{\rm SE}$ first decreases as density increases (note reversed direction of the vertical axis in order to highlight the accuracy of the SE relation at high densities). In the dilute gaseous regime the transport coefficients scale approximately as~\cite{KhrapakPRE04_2021}: $D_{\rm R}\sim \eta_{\rm R}\sim \lambda_{\rm R}\sim (\rho^*)^{-2/3}$. This implied $\alpha_{\rm SE}\sim (\rho^*)^{-4/3}$. If we further neglect a relatively weak temperature dependence of the freezing density in the range of temperatures investigated, we arrive at the scaling $\alpha_{\rm SE}\sim {\mathcal R}^{-4/3}$, shown by the dotted line in Fig.~\ref{FigSE}. This is only a rather crude approximation, since the freezing density actually increases from $\rho_{\rm fr}\simeq 0.88$ at $T^*=0.5$ to $\rho_{\rm fr}\simeq 1.36$ at $T^*=6.0$~\cite{AttiaJCP2022}. Nevertheless, we observe in Fig.~\ref{FigSE} that the numerical data do approach this asymptote as the density decreases.

As the density increases and the freezing line is approached, the data are saturating around a constant value $\alpha_{\rm SE}\simeq 0.15$, shown by a horizontal dashed line in Fig.~\ref{FigSE}. The data points approach this asymptote at ${\mathcal R}\gtrsim 0.6$. This is where the SE relation is applicable and this is where vibrational paradigm of atomic dynamics in liquids holds~\cite{KhrapakPhysRep2024}.  The picture is very similar to that in the LJ fluid, see Fig.~1 from Ref.~\cite{KhrapakPRE10_2021}. Thus, the SE relation without the hydrodynamic diameter works in the dense WCA fluid just as it does in the LJ fluid.

\subsection{Freezing density scaling}

The reduced self-diffusion ($D_{\rm R}$), viscosity ($\eta_R$), and thermal conductivity ($\lambda_{\rm R}$) coefficients as functions of the FDS scaling parameter ${\mathcal R}=\rho/\rho_{\rm fr}$ are shown in Fig. ~\ref{FigTransport}. Symbols correspond to five different isotherms studied in Ref.~\cite{NasrabadJCP2008}. The dashed curves correspond to the freezing density scaling in the LJ fluid, as calculated using the model of Ref.~\cite{BellJPCB2019}. We observe that FDS of transport coefficients in the WCA fluid essentially coincides with that in the LJ fluid. Taking into account that similar FDS operates in the HS fluid~\cite{KhrapakPRE04_2021} and in noble gases~\cite{KhrapakJCP2022_1} indicates that this form of FDS is quasi-universal for sufficiently steep interaction potentials. 

It is observed in Fig.~\ref{FigTransport} that the self-diffusion coefficient decreases monotonically with density, while the shear viscosity and thermal conductivity coefficients increase towards the fluid-solid phase transition. Since in the low density regime the transport coefficients scale as $D_{\rm R}\sim \eta_{\rm R}\sim \lambda_{\rm R}\sim (\rho^*)^{-2/3}$, minima in the reduced viscosity and thermal conductivity coefficients should be expected, just as in other classical fluids~\cite{KhrapakJCP2022_1,KhrapakPoF2022}. These minima correspond to a crossover between different mechanisms of momentum and energy transport and hence to a gas-liquid dynamical crossover (Frenkel line on the phase diagram), which is a topic of considerable recent interest and debate~\cite{BrazhkinPRE2012,BrazhkinPRL2013,BrazhkinJPCB2018,BrykJPCB2018,BellJCP2020,
KhrapakJCP2022}. However, since the available numerical results start from the density $\rho^*=0.3$, the minima are not really observed in Fig.~\ref{FigTransport} and we do not elaborate on this point further.

The FDS that applies to steep interaction potentials such as LJ, WCA, HS cannot be truly universal. For extremely soft interaction potentials such as Coulomb and screened Coulomb (Yukawa) potentials, which are of interest in the plasma-related context~\cite{BausPR1980,FortovUFN,FortovPR,IvlevBook}, the dependence of the transport coefficients on the density is qualitatively similar. The minima in shear viscosity and thermal conductivity occur at $\Gamma/\Gamma_{\rm fr}\simeq 0.05$~\cite{HuangPRR2023,KhrapakPPR2023}, where $\Gamma=Q^2/ak_{\rm B}T$ is the Coulomb coupling parameter, defined as the ratio of the Coulomb energy of interaction at a mean interparticle separation to the kinetic energy, expressed using system temperature. Here $Q$ is the electrical charge, $a=(4\pi\rho/3)^{-1/3}\propto \rho^{-1/3}$ is the Wigner-Seitz radius, and $\Gamma_{\rm fr}$ corresponds to the coupling parameter at the fluid-solid phase transition (freezing). Rewritten in terms of the relative density, the condition for the minima reads $\rho/\rho_{\rm fr} = (\Gamma/\Gamma_{\rm fr})^3\simeq 0.000125$, which is drastically different from $\rho/\rho_{\rm fr}\simeq 0.3$ in the LJ and HS fluids~\cite{KhrapakPRE04_2021,KhrapakPoF2022}. Thus, the shapes of FDS must be different for soft and steep interactions. Where exactly the transition between soft and steep regimes occurs requires further investigation and might be subject of future work.    

Finally, a good representation of the reduced transport coefficients of the dense WCA fluid can be achieved with simple exponential functions of the form 
\begin{align}
D_{\rm R}\simeq 0.3e^{-2.4{\mathcal R}^{2.7}},\label{DR}\\
\eta_{\rm R} = 0.5e^{2.4{\mathcal R}^{2.7}}, \label{etaR}\\ 
\lambda_{\rm R}\simeq 1.2e^{2.15{\mathcal R}^{1.1}}\label{lambdaR}.
\end{align}
These fits are shown by the solid lines in Fig.~\ref{FigTransport}. The agreement with numerical results at $\rho/\rho_{\rm fr}\gtrsim 0.6$ is convincing (but there is still potential for improvement when new data become available). They should not be applied for lower densities, because they are not consistent with the behavior of transport coefficients in this regime.  Note that the fits for the self-diffusion and viscosity coefficients satisfy the SE relation at high densities, $D_{\rm R}\eta_{\rm R}\simeq 0.15$. 
Formulas (\ref{DR}) - (\ref{lambdaR}) are purely empirical fits convenient for practical application. Nevertheless, there is a certain similarity with the excess entropy scaling proposed by Rosenfeld~\cite{RosenfeldPRA1977,RosenfeldJPCM1999}, where the reduced transport coefficients are described as exponential functions of excess entropy. Actually, the FDS scaling parameter ${\mathcal R}$ and the excess entropy are intimately related, as illustrated in Fig.~2 of Ref.~\cite{KhrapakJPCL2022} for the LJ fluid. These relations have not been used in constructing Eqs.~(\ref{DR})-(\ref{lambdaR}). Other functional forms for the dependencies $\eta_{\rm R}({\mathcal R})$ and $D_{\rm R}({\mathcal R})$ have been recently proposed in Ref.~\cite{HeyesJCP2023}.

\section{Conclusion}

Unlike transport properties of dilute gases and solids, transport properties of liquids are much less understood. This is why various approximate relationships and empirical correlations still play important role in describing their transport properties.

This paper has been focused on the freezing density scaling, proposed originally to describe the self-diffusion, viscosity, and thermal conductivity coefficients of the LJ fluid. It has been demonstrated here that the same or very similar FDS applies to the WCA fluid as well. Given the fact that the phase diagrams of the LJ and WCA systems are completely different and taking into account the previous observation that LJ-type FDS is similar to that in the fluid of hard spheres and some liquefied noble gases, our results provide strong indication about quasi-universal character of this FDS. Therefore, we confirm the usefulness of the FDS approach to estimate transport coefficients of dense fluids with steep interatomic interactions. This can be of interest for researchers in chemical physics, (soft) condensed matter, physics of fluids, materials science and beyond.

The authors have no conflicts of interest to disclose.

Data sharing is not applicable to this article as no new data were created or analyzed in this study. 

%\acknowledgments

%\bibliographystyle{aipnum4-1}

\bibliography{SE_Ref}

\end{document}